\documentclass{statsoc}

\usepackage[a4paper]{geometry}
\usepackage{graphicx}
\usepackage[textwidth=8em,textsize=small]{todonotes}
\usepackage{amsmath}
\usepackage{natbib}

\RequirePackage{fix-cm}
\usepackage{graphicx}
\usepackage{natbib}
\usepackage{color}
\usepackage{times}
\usepackage{latexsym}
\usepackage{enumerate}
\usepackage{epsfig}
\usepackage{amsmath}
\usepackage{amssymb}
\usepackage{amsfonts}
\usepackage[english]{babel}
\usepackage{graphicx}
\usepackage{graphics}
\usepackage{psfrag}
\usepackage{amsbsy}
\usepackage{amsmath}
\usepackage{textcomp}
\usepackage{booktabs}
\usepackage{amssymb}
\usepackage{amsmath}
\usepackage{bm}
\usepackage{enumerate}
\usepackage{algorithm}
\usepackage{algorithmic}
\usepackage{multirow}
\usepackage{verbatim}
\usepackage{caption}
\usepackage{subcaption}
\usepackage{tabularx}
\usepackage{listings}

\usepackage{mathptmx}       
\usepackage{helvet}         
\usepackage{courier}        
\usepackage{type1cm}        
\usepackage{makeidx}         
\usepackage{graphicx}        
\usepackage{multicol}        
\usepackage[bottom]{footmisc}

\usepackage{amsmath}
\usepackage{amsfonts}

\usepackage{amsmath}
\usepackage{amsfonts}

\newcommand{\vecy}{\mathbf{y}}

\newcommand{\vecmu}{\mbox{\boldmath$\mu$}}

\newcommand{\matsig}{\bm{\Sigma}}

\newcommand{\varthet}{\boldsymbol{\vartheta}}

\DeclareMathOperator*{\argmax}{arg\,max}

\def\be{\mathbf{e}}
\def\bx{\mathbf{x}}

\def\bm{\mathbf{m}}

\def\bX{\mathbf{X}}

\def\bI{\mathbf{I}}

\def\bU{\mathbf{U}}

\def\b0{\mathbf{0}}
\def\b1{\mathbf{1}}

\def\cN{\mathcal{N}}

\def\bmu{\mbox{\boldmath $\mu$}}

\def\btheta{\mbox{\boldmath $\theta$}}

\def\bSigma{\mbox{\boldmath $\Sigma$}}

\def\bLambda{\mbox{\boldmath $\Lambda$}}

\def\bPsi{\mbox{\boldmath $\Psi$}}

\title[Auto Diff in Mixtures]{  Automatic Differentiation in Mixture Models}
\author[Author 1 {\it et al.}]{Siva Rajesh Kasa Vaibhav Rajan}
\address{DISA, School of Computing, NUS}
\email{kasa@u.nus.edu}
\address{Reproducible code available @: \textit{upon email request}}

\begin{document}
\begin{abstract}

In this article, we discuss two specific classes of models - Gaussian Mixture Copula models and Mixture of Factor Analyzers - and the advantages of doing inference with gradient descent using automatic differentiation. 
Gaussian mixture models are a popular class of clustering methods, that offers a principled statistical approach to clustering.  However, the underlying assumption, that every
mixing component is normally distributed, can often be too
rigid for several real life datasets. In order to to relax the assumption about
the normality of mixing components,
a new class of parametric mixture models that are based on
Copula functions - Gaussian Mixuture Copula Models were introduced.  Estimating the parameters
of the proposed Gaussian Mixture Copula Model (GMCM) through maximum likelihood has been intractable due to the positive semi-positive-definite constraints on the variance-covariance matrices. Previous attempts were limited to maximizing a proxy-likelihood which can be maximized using EM algorithm. These existing methods, even though easier to implement, does not guarantee any convergence nor monotonic increase of the GMCM Likelihood.  In this paper, we use automatic differentiation tools to maximize the exact likelihood of GMCM, at the same time avoiding any constraint equations or Lagrange multipliers. We show how our method leads a monotonic increase in likelihood and converges to a (local) optimum value of likelihood. 

In this paper, we also show how Automatic Differentiation can be used for inference with Mixture of Factor Analyzers and advantages of doing so. We also discuss how this method also has all the properties such as monotonic increase in likelihood and convergence to a local optimum. 

Note that our work is also applicable to special cases of these two models - for e.g. Simple Copula models, Factor Analyzer model, etc.

\keywords{Constrained Optimization, \and Gaussian Mixture Copula Models,  \and Mixture of Factor Analyzers, \and Automatic Differentiation}

\end{abstract}

\section{Introduction}

\subsection{Gaussian Mixture Copulas}

Clustering is a fundamental problem in data mining with a large number of applications \cite{jain2010data}.
Innumerable clustering methods have been developed, that often differ in input datatypes, application domains as well as methodogy \cite{aggarwal2013data}.
Model-based clustering is a popular class of clustering methods, that offers a principled statistical approach to clustering.
The clusters can be interpreted through the lens of the underlying distributional assumptions of the model.
Assuming a generative probabilistic model, such as a finite mixture model,
allows us to use statistical inference methods to solve unsupervised pattern recognition problems such as cluster and outlier detection.


Gaussian mixture models are perhaps the most widely used model-based clustering method.
Both assumptions of normal distribution as well as same distributions in all components are often violated in real data.
Model-based approaches can leverage the flexible framework of copulas that provides a modular parameterization of multivariate distributions -- arbitrary marginals independent of dependency models from copula families which can model a wide variety of linear and non–linear dependencies.
Copula mixture models \cite{fujimaki2011online} as well as mixture copulas \cite{compstat} have been used for clustering.  
The former uses copula-based distributions in a mixture model whereas the latter use mixture models to define a copula family.


The Gaussian mixture copula model (GMCM) combines the modeling strengths of a copula-based approach and Gaussian mixture models.
It allows flexible dependency modeling, especially of non-Gaussian data, and can model many kinds of multi-modal dependencies, notably asymmetric and tail dependencies.
Copula models, including GMCM, are rank-based which makes them invariant to monotone increasing marginal transformations.


\subsection{Mixture of Factor Analyzers}

Finite mixture distributions are well-known in statistical modeling because they bring the flexibility of non-parametric models while preserving the strong mathematical proprieties of parametric models. In this paper, we extend automatic differentiation to mixture factor analyzers (MFA), a classic high-dimensional modeling tool.  According to MFA model, we assume that a sample of observations has been drawn from different populations, whose latent structures are modeled using low-dimensional individual factors. The aim is to decompose the sample into its mixture components, which are usually modeled using a multivariate Gaussian distribution, and to estimate parameters. The assumption of component-wise normality, besides its convenient expression in a closed-form  for multi-variate distributions, also allows to employ the EM algorithm for the ML estimation of the parameters. However, in the recent past, the rise in automatic differentiation tools available allows us to do inference using a gradient based approach (Auto-MFA).

\subsection{Our Contribution}
Previous parameter inference methods for GMCM either use a proxy likelihood (\cite{bilgrau2016gmcm}  \cite{compstat})or do use constraint equations for positive-semi-definite co-variance matrices \cite{tewari2011parametric} which are computationally intractable.
In this paper, we present a new methodology, called Auto-GMCM, whereby we tackle these two problems using Automatic Differentiation tools. Our method, which maximizes the exact GMCM likelihood, does not use any constraints nor any Lagrange multipliers, can be scaled to any number of dimensions and can be implemented easily using popular data-analysis software such as R and Python. Moreover, we show how our method results in monotonic increase and convergence of the likelihood to the local optimum, which is absent in the existing methods.

For MFAs, a) we show how our Auto-MFA maximizes the likelihood better compared to EM-based MFA, because of availability of second order derivatives (such as hessian matrix) b) we show our method Auto-MFA is robust to high-dimensional settings (n $\le$ p) whereas EM-based MFA fails.

The rest of the paper is organized as follows: First, we give a brief introduction of Automatic Differentiation and its advantages. Next, we discuss the details of GMCM and  parameter inference using our method, Auto-GMCM. Later, we compare our method with the the existing method on simulated data and illustrate the key differences.  Thereafter, we do a similar study on MFAs and parameter inference using Auto-MFA, and discuss the advantages of doing inference using Auto-MFA on a real dataset. 

\section{Brief Introduction of Automatic Differentiators -  \cite{baydin2015automatic}}

Automatic Differentiation is the workhorse of optimization routines used in todday's machine learning. While it is ease to see the difference between Automatic Differentiation (AD) vs Numerical Differentiation (ND)  (which is marred by approximation errors), the subtle difference between AD vs Symbolic Differentiation (SDs) is not easy to appreciate. 

The key idea is this: In symbolic differentiation, we first naively evaluate the complete expression and then differentiate a complex expression using sum rule, product rule, etc. However, it may so happen that we are indeed repeatedly evaluating a same expression multiple times. For e.g.

\begin{equation}
\begin{aligned}
\frac{d}{dx} \left(f(x) + g(x)\right) &\leadsto \frac{d}{dx} f(x) + \frac{d}{dx} g(x)\\
\frac{d}{dx} \left(f(x)\,g(x)\right) &\leadsto \left(\frac{d}{dx} f(x)\right) g(x) + f(x) \left(\frac{d}{dx} g(x)\right)\; .
\end{aligned}
\label{EquationMultiplicationRule}
\end{equation}

Consider a function $h(x)=f(x)g(x)$ and the multiplication rule in the above equation. Since $h$ is a product, $h(x)$ and $\frac{d}{dx}h(x)$ have some common components, namely $f(x)$ and $g(x)$. Note also that on the right hand side, $f(x)$ and $\frac{d}{dx}f(x)$ appear separately. If we just proceeded to symbolically differentiate $f(x)$ and plugged its derivative into the appropriate place, we would have nested duplications of any computation that appears in common between $f(x)$ and $\frac{d}{dx}f(x)$. Hence, careless symbolic differentiation can easily produce exponentially large symbolic expressions which take correspondingly long to evaluate. This problem is known as \textbf{expression swell}.

When we are concerned with the accurate numerical evaluation of derivatives and not so much with their actual symbolic form, it is in principle possible to significantly simplify computations by storing only the values of intermediate sub-expressions in memory. Moreover, for further efficiency, we can interleave as much as possible the differentiation and simplification steps. This interleaving idea forms the basis of AD and provides an account of its simplest form: \textbf{apply symbolic differentiation at the elementary operation level and keep intermediate numerical results, in lockstep with the evaluation of the main function.} 

To illustrate the problem, consider the following example: Iterations of the logistic map $l_{n+1}=4l_n (1-l_n)$, $l_1=x$ and the corresponding derivatives of $l_n$ with respect to $x$, illustrating expression swell. The table \ref{TableExpressionSwell} clearly shows that the number of repetitive evaluations increase with $n$.

\begin{center}
  \centering
  \captionof{table}{Iterations of the logistic map $l_{n+1}=4l_n (1-l_n)$, $l_1=x$ and the corresponding derivatives of $l_n$ with respect to $x$, illustrating expression swell. Taken from \cite{baydin2015automatic} }
  \label{TableExpressionSwell}
  \renewcommand{\arraystretch}{1.2}
  
  {\small
  \begin{tabularx}{\columnwidth}{@{}lXp{4cm}XX@{}}
    \toprule
    $n$ & $l_n$ & $\frac{d}{dx}l_n$ & $\frac{d}{dx}l_n$ (Simplified form)\\
    \addlinespace
    \midrule
    1 & $x$ & $1$ & $1$\\
    \addlinespace
    2 & $4x(1 - x)$ & $4(1 - x) -4x$ & $4 - 8x$\\
    \addlinespace
    3 & $16x(1 - x)(1 - 2 x)^2$ & $16(1 - x)(1 - 2 x)^2 - 16x(1 - 2 x)^2 - 64x(1 - x)(1 - 2 x)$ & $16 (1 - 10 x + 24 x^2 - 16 x^3)$\\
    \addlinespace
    4 & $64x(1 - x)(1 - 2 x)^2$ $(1 - 8 x + 8 x^2)^2$ & $128x(1 - x)(-8 + 16 x)(1 - 2 x)^2 (1 - 8 x + 8 x^2) + 64 (1 - x)(1 - 2 x)^2  (1 - 8 x + 8 x^2)^2 - 64x(1 - 2 x)^2 (1 - 8 x + 8 x^2)^2 - 256x(1 - x)(1 - 2 x)(1 - 8 x + 8 x^2)^2$ & $64 (1 - 42 x + 504 x^2 - 2640 x^3 + 7040 x^4 - 9984 x^5 + 7168 x^6 - 2048 x^7)$\\
    \bottomrule
  \end{tabularx}}
\end{center}

A rudimentary solution to counter expression swell would be: for e.g.: $x(1-x)$ occurs many times in the derivative. So, while computing the $\frac{dl}{dx}$ say at $x = 0.5$, we can compute this value of $x(1-x) = 0.5*0.5 = 0.25$ once and use it whenever we encounter.What AD does actually is that rather than first getting the expression of $l_{n+1}$ entirely in terms of x and then differentiating, we can do the following: 
The derivative of $l_{n+1} = 4 l_{n} (1- l_{n})$ can be found using the chain rule $\frac{dl_{n+1}}{dl_n}\frac{dl_{n}}{dl_{n-1}}\dots \frac{dl_{1}}{dl_x}$ which simplfies to $4(1-l_{n} - l_{n})4(1-l_{n-1} - l_{n-1})\dots4(1-x - x)$. 

The thing to note is that evaluating the AD way is computationally linear wrt $n$ (because we add only one $(1-l_n - l_n)$ for each increase by 1) whereas using the SD way is exponential (as evident in the table above). This linear time complexity is achieved due to carry-over of the derivatives at each step, rather than evaluating the derivative at the end and substituting the value of x. This is the crux of AD. Of course, there are other important differences between AD vs SD but they are not considered here. AD tools are available in many software packages. Here, we give an example solution  using the \texttt{Autograd} package in Python. 

    \begin{lstlisting}
from autograd import grad
def my_func(x,n):
    p = x
    y = x*(1-x)
    for i in range(n):
        y = y*(1-y)
        
    return y
grad_func = grad(my_func)
grad_func(0.5,4)
    \end{lstlisting}
    
Also, doing the differentiation using the SD way is suitable only when the function is expressed in a closed mathematical form such as polynomials, trigonometric functions, exponential functions, etc. However, if the function is a computer program with control structures such as \texttt{for, if, while} - e.g. the above Python function -, in such cases, we cannot use SD, as the function is not in a closed-form. In such cases only AD and ND are suitable. 

\subsection{Runtime comparison of Automatic and Symbolic Differentiation}

Consider the following recursive expressions: 
$l_0 = \frac{1}{1+e^x}$, $l_1 = \frac{1}{1+e^{l_0}} $, ....., $l_{n} = \frac{1}{1+e^{l_{n-1}}} $

We evaluate the derivative of $l_{n}$ wrt x and compare the runtime in Mathematica (SD) vs Python (AD) for various values of $n$. As $n$ increases, it is expected that runtime also increases. However, it can be seen from the results in Table \ref{Runtime}  that runtime increases linearly for Python (AD) whereas it increases exponentially for Mathematica (SD).

\begin{center}
\captionof{table}{Average runtime (over 1000 runs) }
\label{Runtime}
\begin{tabular}{ |c|c|c| } 

 \hline
 n & \text{Python}  & \text{Mathematica} \\ \hline
1 & 0.00013 & 0.00000 \\ 
5 & 0.00030 & 0.00005 \\ 
10 & 0.00051 & 0.00023 \\
50 & 0.00293 & 0.00437 \\
100 & 0.00433 & 0.15625 \\
200 & 0.00917 & 1.45364 \\
 \hline
\end{tabular}
\end{center}

\section{Copulas in Mixture Models}
Since copulas provide a flexible characterization of multivariate distributions, they have been used in mixture models by several authors, e.g., 
\cite{fujimaki2011online},
\cite{tekumalla2017vine}.
None of these specifically address the problems of clustering high-dimensional data.
Vine copulas, that are hierarchical collections of bivariate copulas, can scale to moderately high dimensions but at the cost of exponentially increasing complexity for model selection and estimation \cite{muller2018representing}. 
See \cite{elidan2013copulas} for a comparison of copulas with machine learning models including a discussion on fitting copulas to high-dimensional data and \cite{joe2014dependence} for a comprehensive treatment of copulas.

The Gaussian Mixture Copula model (GMCM) was proposed by \cite{tewari2011parametric}.
Unlike mixtures of copulas, GMCM is a copula family where the (latent) copula density follows a Gaussian mixture model (the following section has details).
This has considerable advantages for copula-based clustering since
clusters can be inferred directly from the dependencies 
obviating the need for marginal parameter estimation.
This was leveraged for clustering by \cite{compstat} who designed an EM-based algorithm for GMCM parameter estimation.
A mixed EM and Gibbs Sampling based approach, for clustering was designed by 
\cite{rajan2016dependency} to fit real and ordinal data.
\cite{li2011reproducibility} studied a specific case of GMCM to design a meta-analysis method called reproducibility analysis, to verify the  reliability and consistency of multiple high-throughput genomic experiments (that yield high dimensional data).
Computational and statistical hurdles in GMCM parameter estimation were discussed and alternate workarounds involving numerical optimization methods  were suggested in 
\cite{bilgrau2016gmcm}.

\subsection{Gaussian Mixture Copula Model}
\label{model}

A $p$--dimensional copula is a multivariate distribution function $C: [0,1]^p \mapsto [0,1]$. 
A theorem by \cite{Sklar} proves that copulas can uniquely characterize continuous joint distributions:
for every joint distribution with continuous marginals, $F(Y_1, \ldots, Y_p)$, there exists a 
unique copula function such that $F(Y_1, \ldots, Y_p) = C(F_1(Y_1), \ldots , F_p(Y_p) )$
as well as the converse.
Parametric copula families are typically defined on uniform random variables obtained through CDF transformations from the marginals.  
In a Gaussian Mixture Copula Model (GMCM), the dependence is obtained from a Gaussian Mixture (GMM).

Consider a $p$-dimensional, $G$-component Gaussian Mixture Model, $\mathcal{G}(\varthet)$, parameterized by $\varthet= (\pi_1,...\pi_G, \vecmu_1,...,\vecmu_G, \matsig_1,...,\matsig_G)$, representing mixing proportions ($\pi_g>0$, with $\sum_{g=1}^G\pi_g=1$), mean vectors ($\vecmu_g$) and covariance matrices ($\matsig_g$) for components $g=1,\ldots,G$.
Let $\Psi_j(\varthet)$ 
and $\psi_j(\varthet)$ denote the $j^{th}$ marginal CDF and PDF respectively, of $\mathcal{G}(\varthet)$; $\Phi_j$ and $\phi_j$ denote the $j^{th}$ marginal CDF and PDF, respectively of the multivariate normal distribution with PDF $\phi$.

{\bf GMCM} assumes a generative process for $n$ instances of the observed data, $\mathbf{X} = [x_{ij}]_{n \times p} = (\mathbf{X}_1,\ldots,\mathbf{X}_p)$, with arbitrary marginal CDFs $F_j$, $j = 1,\ldots,p$, specified by\footnote[1]{single subscript denotes column index here}:
\begin{equation}\label{GMCM}
\begin{split}
\mathbf{X}_j = F_j^{-1}(\mathbf{U}_j);\,\, \mathbf{U}_j = \Psi_j(\mathbf{Y}_j);\,\,  \mathbf{Y} \sim \mathcal{G}(\varthet), \\ \forall j \in \{1,\ldots,p\}.
\end{split}
\end{equation}
Note that the variables $\mathbf{U} = [u_{ij}]_{n \times p} = (\mathbf{U}_1,\ldots,\mathbf{U}_p)$ where $\mathbf{U}_j = F_j(\mathbf{X}_j) = \Psi_j(\mathbf{Y}_j)$ are probability integral transforms and have uniformly distributed marginals.
The PDF $\mathcal{C}$ of $\mathbf{U}$ is the {\it copula density} of $\mathcal{G}$ \cite{joe2014dependence}, given by:
\begin{equation}\label{GMC}
\mathcal{C}(\mathbf{U};\varthet) = 
\frac{\psi (\Psi_j^{-1} (\mathbf{U}))}{\prod_{j=1}^p \psi_j( \Psi_j^{-1} (\mathbf{U}_j))} 
\end{equation}
Thus, the likelihood of $n$ i.i.d. samples from GMCM is:
\begin{equation}\label{exactGMCM}
\mathcal{C} = \prod_{i=1}^n \frac{\sum_{g=1}^G \pi_g\phi(\vecy_i\mid\vecmu_g,\matsig_g)}{\prod_{j=1}^p \psi_j(y_{ij} \mid\vecmu_g,\matsig_g))}
\end{equation}

For clustering, GMCM can be used to obtain cluster labels $l \in {1,\ldots,G}$ through a semiparametric MAP estimate $\argmax_l P(l = g| \varthet, \mathbf{X})$ without estimating the marginal parameters \cite{compstat}.
A special case of GMCM in which $\mathcal{G}$ has 2 components, $\matsig_1$ is the identity matrix, $\matsig_2$ has an equi-covariance structure, $\vecmu_1 = \mathbf{0}_{1 \times p}$ and $\vecmu_2 = (\mu, \ldots, \mu)_{1 \times p}$ has equal and positive values, is used for reproducibility analysis \cite{li2011reproducibility,bilgrau2016gmcm}.

\subsection{GMCM Parameter Inference, under existing methods}
\label{inference}

Maximizing the exact likelihood of GMCM $\mathcal{C}$ is intractable using EM and, in practice, the pseudo-likelihood (PEM) below is used:
\begin{equation}\label{pseudo}
\mathcal{L} = \prod_{i=1}^n \sum_{g=1}^G \pi_g\phi(\vecy_i\mid\vecmu_g,\matsig_g)
\end{equation}
Genest, Ghoudi and Rivest \cite{genest1995semiparametric} study the properties of estimates based on the pseudo-likelihood and show that for continuous-valued marginals, the estimator is consistent and asymptotically normal.
They have also been used for obtaining parameter estimates of the Gaussian copula \cite{hoff2007extending}.
However, even obtaining a Maximum Likelihood (ML) estimate through the pseudo-likelihood,
$\argmax_{\varthet} \mathcal{L} (\mathbf{U})$, poses challenges for GMCM that are detailed in \cite{bilgrau2016gmcm}.

The main challenge is due to the inverse CDF $\Psi_j^{-1}$ in equation \ref{GMC} through which we obtain $\vecy_i$, in equation \ref{pseudo}.
This inverse CDF has no closed-form expression. Several different ways have been proposed to tackle this lack of closed-form expression. A grid search and linear interpolation was suggested by \cite{bilgrau2016gmcm}. The drawback of this method is that it cannot be scaled to high dimensions (order of 100's). \cite{compstat} overcame this scalability problem by suggesting an approximation to compute the inverse CDF . In all the three previous papers, i.e. \cite{bilgrau2016gmcm,tewari2011parametric,compstat} , pseudo EM approach (PEM) is used, which iteratively alternates between estimating $y_{ij} = \Psi_j^{-1}(\hat{\mathbf{U}_j}, \varthet)$ and updating $\varthet$ by E and M steps. 
 In addition, \cite{tewari2011parametric} also proposes an alternative gradient based approach, using constraints on positive-semi-definiteness of covariance matrices and Lagrange multipliers for mixture components $\pi_g$. However, the paper also identifies that it is cumbersome to evaluate the gradients, while simultaneously maintaining the positive semi-definiteness constraints on each of the co-variance matrices $\matsig_g$

\section{Reasons for using EM based methods previously}
Previous techniques relied on maximizing the proxy-likelihood mentioned in equation \ref{pseudo}. Typically Expectation Maximization (EM) is used to solve the problem of maximizing the likelihood of $\mathcal{L}(X|\theta)$ where $X$ is the observed data and $\theta$ is the parameter vector. Any introductory material on EM starts off by showing how EM simplifies the maximization in the case of Gaussian Mixture Models (E.g: Chapter 9, PRML, C. Bishop). EM and its variants are widely used in the problems of latent/missing/unobserved data. There are a few reasons why EM has been so popular. Some of them, in the context of GMCMs, are discussed below:

\begin{itemize}
    \item Solving the $\text{argmax}_{\theta}\mathcal{C}(X|\theta)$ problem is typically intractable if one tries to evaluate the gradients and set them to zero. Using EM, one tries to reformulate the above problem using its lower bound and tries to maximize the lower bound. It is expected that maximizing this lower bound is tractable and easier. 
    
\item In the case of GMMs,  the parameter vector contains the covariance matrices $\Sigma_g$ for each component $g$. 
\item \textit{Problem 1:} Evaluating derivatives wrt to matrices is inconvenient i.e. $\frac{\partial L }{\partial \Sigma_g}$ is almost intractable. Using EM, we get simplify the problem into a sequence of simpler optimization problems where the gradients are easy to obtain and immediately set to zero, i.e. no need for any gradient descent approach.  

\item \textit{Problem 2:} Moreover, without EM, maximizing $\mathcal{L}(X|\theta)$ with $\Sigma_g$ would require complicated constraints on the positive semi definiteness (PSD) of the estimate $\hat{\mathcal{L}(X|\theta)}$ we are trying to obtain. However, EM eliminates the need for any such constraints by nature of its formulation.

\item \textit{Problem 3:} To ensure that sum of proportions of components add up to one, typically a Lagrange multiplier is used. However, in the case of GMCM, using a Lagrange multiplier will only further aggravate the complexity and intractability of the problem. 
 
\end{itemize}

\section{Auto-GMCM: An EM-free Inference Algorithm}

Our new method, called Auto-GMCM, avoids the problems associated with previous methods EM in the following ways. 

With the proliferation of automatic differentiating (AD) tools, an alternate numerical optimization route, to maximize the exact GMCM likelihood,   $\mathcal{C}(X|\theta)$ mentioned in equation \ref{exactGMCM} is possible. 

To tackle \textit{Problem 2} instead of gradients with respect to $\Sigma_g$, we get the gradients wrt $U_g$, where $\Sigma_g = U_gU^{T}_g$. We first initialize the values of $U_g$ be identity matrices. Thereafter, we keep adding the gradients to the previous values of $U_g$, i.e. $U_g := U_g + \alpha * \frac{\partial \mathcal{L}}{\partial U_g}$. Here, $\alpha$ is the learning rate. The reasoning is as follows: If the gradients are evaluated with respect to $\Sigma_g$ directly, there is no guarantee that updated $\Sigma_g = \Sigma_g + \frac{\partial \mathcal{C}}{\partial \Sigma_g}$ will still remain PSD. However, if we gradients are evaluated with respect to $U_g$, no matter what the updated matrix $U_g$ is, by construction $\Sigma_g$ always remains PSD. Alternately, one could use a Cholesky decomposition for $\Sigma_g$

\textit{Problem 1} is solved inherently by use of ADs because we need not write/code out any gradient manually. What's more is that because we can evaluate the second-order Hessian matrix using ADs, we can implement faster convergence optimization methods like Newtons method, not just gradient descent.

\textit{Problem 3} is solved considering the log proportions trick, using the logsumexp trick \cite{robert2014machine}. We start with unbounded $\alpha_g$'s as the log-proportions i.e.  $\log \pi_g = \alpha_g - {\log(\sum_i e^{\alpha_i})}$.  Note that, we need not impose any constraints on $\alpha_g$ as final computation of $\pi_g$ automatically leads to normalization, because $\pi_g = \frac{\alpha_g}{\sum_i e^{\alpha_i}}$. Therefore, we can update $\alpha_g := \alpha_g + \frac{\partial C}{\partial \alpha_g}$ without any further need for Lagrange multipliers. 

Once the gradients are obtained, we use the grid search method of \cite{bilgrau2016gmcm} to obtain the updated $y_i$'s. 
\subsection{Illustration and Results}
We simulate a 3 - component, 2 - dimensional, wrapped mixture model for the purpose of illustration. Wrapped mixtures are not just linear transformations of our latent Gaussian model but also a radial distortion is applied as a transformation on the Gaussian data \cite{johnson2016composing}. 

\begin{figure}[h!]
 \centering
\begin{subfigure}{0.85\textwidth}
\includegraphics[width=0.9\linewidth, height=5cm]{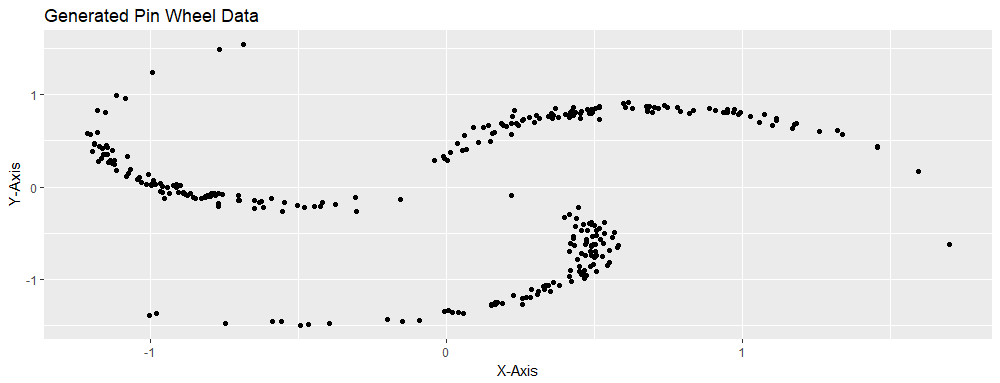} 
\caption{Iteration 1}
\label{fig:subim1}
\end{subfigure}
\begin{subfigure}{0.85\textwidth}
\includegraphics[width=0.9\linewidth, height=5cm]{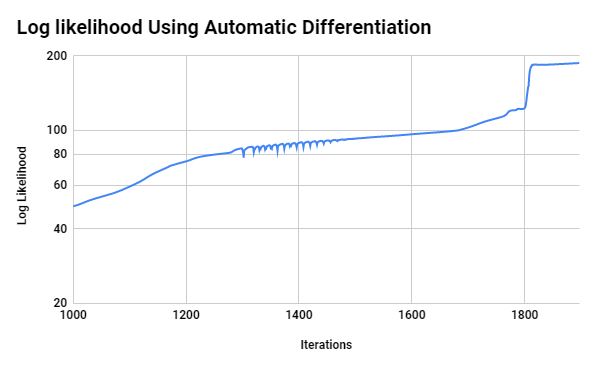}
\caption{Auto-GMCM}
\label{Auto-GMCM}
\end{subfigure} 

\begin{subfigure}{0.85\textwidth}
\includegraphics[width=0.9\linewidth, height=5cm]{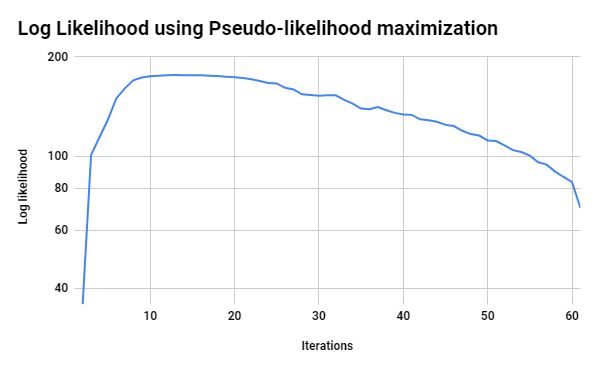} 
\caption{Pseudo-Likelihood maximization}
\label{Pseudo-GMCM}
\end{subfigure}
\\

\caption{MLE OF GMCM existing method vs Auto-GMCM}
\label{fig:image2}
\end{figure}

As indicated in figure \ref{fig:image2} b, the likelihood using Auto-GMCM increases monotonically and converges to an optimum value. Moreover, the implementation of Auto-GMCM is pretty straightforward using popular tools such as R and Python (Code can be provided by contacting the authors). 

The PEM algorithm, as illustrated in figure \ref{fig:image2} c, is implemented using GMCM package in R. Both our Auto-GMCM and PEM had the same initialization. Nevertheless, PEM quickly reaches the local maxima but it does not converge nor the likelihood increases monotonically. On the other hand, Auto-GMCM is slow to converge (depends on the learning rate $\alpha$ in gradient descent). Nevertheless, Auto-GMCM increases the likelihood almost monotonically and converges to a higher value compared to PEM approach.

\section{Mixture of Factor Analyzers}
For multivariate data of a continuous nature, a major chunk of past literature has focussed on the use of multivariate normal components, because of their closed form expression. 
Within the Gaussian Mixture Model (GMM) -based approach to density estimation and clustering, the density of the $p$-dimensional random variable $\bX$ of interest is modelled as a mixture of a number, say $G$, of multivariate normal densities in some unknown proportions $\pi_1,\ldots \pi_G$. That is, each data point is taken to be a realization of the mixture probability density function,
\begin{equation}
f(\bx;\btheta)=\sum_{g=1}^G \pi_g \phi_p(\bx;\bmu_g,\bSigma_g)\label{mixt-gaussian}
\end{equation}
where $\phi_d(\bx;\mu,\bSigma)$ denotes the $d$-variate normal density function with mean $\bmu$ and covariance matrix $\bSigma$. Here the vector $\btheta_{GM}(p,G)$ of unknown parameters consists of the $(G-1)$ mixing proportions $\pi_g$, the $G \times p$ elements of the component means $\mu_g$, and the ${1 \over 2} G p (p+1)  $ distinct elements of the component-covariance matrices $\bSigma_g$.
Therefore, the $G$-component normal mixture model (\ref{mixt-gaussian}) with unrestricted component-covariance matrices is a highly parameterized  model. We need some way to  parsimoniously specify the matrices  $\bSigma_g$, because they requires $O(p^2)$ parameters.
Among the various proposals for dimensionality reduction, we demonstrate our method here by considering Mixtures of Factor Analyzers (MFA), proposed by \cite{Ghah:Hilt:1997} and developed further that by \cite{McLa:Peel:fini:2000}.

This MFA model allows to explain data by explicitly modeling correlations between variables in multivariate observations. It postulates a finite mixture of linear sub-models for the distribution of the full observation vector $\bX$, given the (unobservable) factors $\bU$. That is one can provide a local dimensionality reduction method by assuming that the distribution of the observation  $\bX_i$  can be given as
\begin{equation}
\bX_i=\bmu_g+\bLambda_g\bU_{ig}+\be_{ig} \quad\textrm{with probability }\quad \pi_g \: (g=1,\ldots,G) \quad \textrm{for} \,\, i=1,\ldots,n, \label{factor_an}
\end{equation}
where $\bLambda_g$ is a $p \times q$ matrix of \textit{factor loadings}, the  \textit{factors} $\bU_{1g},\ldots, \bU_{ng}$ are $\cN(\mathbf{0},\bI_q)$ distributed independently of the  \textit{errors} $\be_{ig}$, which are  independently $\cN(\mathbf{0},\bPsi_g)$  distributed, and $\bPsi_g$ is a $p \times p$ diagonal matrix $(g=1,\ldots,G)$. We suppose that $q<p$, which means that  $q$ unobservable factors are jointly explaining the $p$ observable features of the statistical units.
Under these assumptions, the mixture of factor analyzers model is given by (\ref{mixt-gaussian}), where the $g$-th component-covariance matrix $\bSigma_g$ has the form
\begin{equation}
\bSigma_g=\bLambda_g \bLambda'_g+\bPsi_g \quad (g=1,\ldots,G). \label{Sigmag}
\end{equation}

Note that this model is a superset of Gaussian Mixture Model and single latent factor analyzer model, essentially bridging dimensionality reduction and mixture models. \textbf{Therefore, our method can be applied to special cases such as  - a) probabilistic principal component analysis (PPCA) model \citep{tipping1999probabilistic} which is a special case of the factor analysis model because it assumes that the distribution of the error term is isotropic and b) parsimonious Gaussian mixture model proposed by \cite{McNi:Murp:Pars:2008}.} 

Given, $x_1,x_2,..,x_n$ i.i.d observations, the likelihood of the MFA model is given by 

\begin{align}
   \mathcal{L}(x) =  \sum_{i}^{n} \sum_{g=1}^{G} \frac{\pi_g}{(2\pi)^{p/2}|\Lambda_g\Lambda_g^{'} + \Psi_g|^{\frac{1}{2}}} \text{x} \exp\{-\frac{1}{2} (\bx_i - \bmu_g)^{'} (\Lambda_g\Lambda_g^{'} + \Psi_g)^{-1} (\bx_i - \bmu_g) \}
    \label{likelihood}
\end{align}

Maximizing the above likelihood (eqn \ref{likelihood}) has been done by Expectation Maximization (EM) \citep{Ghah:Hilt:1997,McLa:Peel:fini:2000} or its variants such as Alternate Expectation Conditional Maximization (AECM) algorithm \citep{McNi:Murp:Pars:2008}. There are two advantages in using algorithms based on EM or its variants - a) Positive Semi-Definiteness (PSD) of the estimates of $\bSigma_g$ will be maintained by construction b) the estimates of mixture-proportions $\pi_g$ will add up to 1 without any need for Lagrange Multipliers.  


\subsection{Auto-MFA: Our Method and results}

Automatic Differentiators can provide the exact gradients of equation \ref{likelihood}. Please refer to  \citep{baydin2018automatic} for more details on the Automatic Differentiators. In order to preserve the PSD of $\bSigma_g$ and constraint on $\sum \pi_g = 1$, we use the following simple tricks \citep{maclaurin2016modeling}: 

\begin{enumerate}
    \item We write the variance term of error $\bPsi_g = \psi_g \psi_g^{T}$ where $\psi$ is a diagonal matrix that contains non-zero terms as $\sqrt{\Psi_{g_{ii}} }$. Now we take the gradients of the loglikelihood $\mathcal{L}(x)$ with respect to $\psi_g$ and update according $\psi_g := \psi_g + \alpha \frac{\partial \mathcal{L}}{\partial \psi_g}$. Here $\alpha$ is the learning rate. This way the PSD of $\bSigma_g = \bLambda_g \bLambda'_g+\bPsi_g$ is always preserved through out all the updated values.
    \item The constraint $\sum \pi_g =1$ is tackled considering the log proportions trick, using the logsumexp trick \citep{robert2014machine}. We start with unbounded $\alpha_g$'s as the log-proportions i.e.  $\log \pi_g = \alpha_g - {\log(\sum_i e^{\alpha_i})}$.  Note that, we need not impose any constraints on $\alpha_g$ as final computation of $\pi_g$ automatically leads to normalization, because $\pi_g = \frac{\alpha_g}{\sum_i e^{\alpha_i}}$. Therefore, we can update $\alpha_g := \alpha_g + \frac{\partial l}{\partial \alpha_g}$ without any further need for Lagrange multipliers.

\end{enumerate}

\subsection{Data and Results}
\begin{figure}[!hb]
    \centering
    \includegraphics[width=0.85\textwidth]{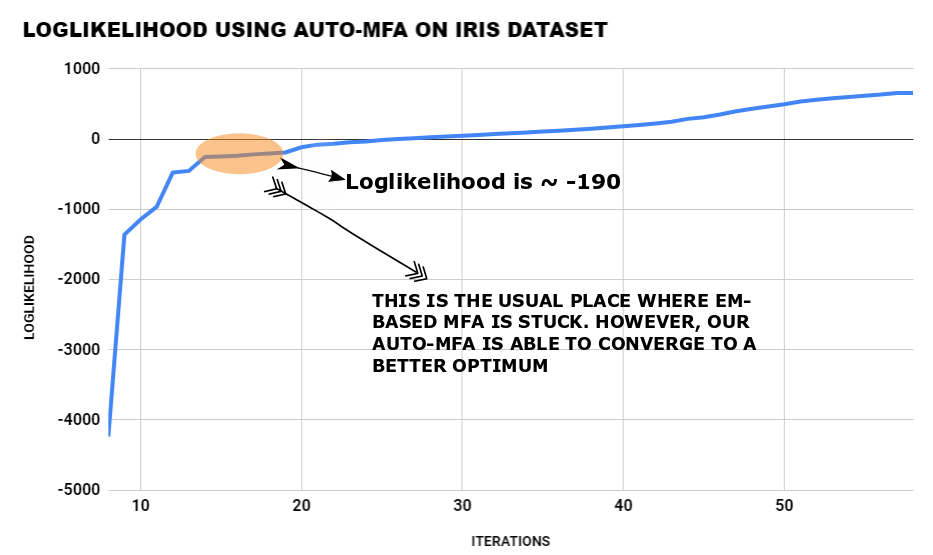}
    \caption{On the IRIS dataset, Auto-MFA converges to a better optimum compared an EM-based MFA}
    \label{fig:my_label}
\end{figure}
We considered the IRIS dataset from \cite{fisher1936use}. It contains sepal length, sepal width, petal length, petal width and class of the 150 different flowers. For the MFA inference using EM, we use the \texttt{EMMixmfa} and \texttt{FactMixAnalysis} R packages. \texttt{EMMixmfa} package contains an option for implementing the case considered in our equation \ref{likelihood} i.e. all the covariance matrices $\bSigma_{b}$ and $\Psi_g$ are different for each of the component. Moreover, we consider the most generalized case where $\Psi_g$ is not isotropic.

We use \texttt{iris} dataset to compare our algorithm Auto-MFA vs EM-based MFA. Because we have access to second order information such as the hessian matrix, we can use Newton-CG method in Auto-MFA. We run both the algorithms with ten different random initializations. The best of loglikelihood of Auto-MFA was 656 while that of EM-MFA is -180 (Figure \ref{fig:my_label}).  However, as expected, the average runtime execution was slower using Auto-MFA (20 seconds) compared to EM-based MFA (2 seconds). This is because run time of gradient-based approach usually depend on learning rate and lower learning rate can lead to higher run time. Moreover, higher runtime can also be attributed to computing the second-order derivatives (hessian matrix). The trade-off for higher runtime is convergence to a better local optimum as evident in Figure \ref{fig:my_label}. Further more, the lower learning rate ensures that the likelihood increases monotonically at every step.     

Moreover, Auto-MFA can even work on high dimensional data (n $\le$ p) whereas a traditional EM-based MFA fails in this case because EM-based MFA involves matrix inversion steps and in high-dimensional settings, this matrix is not full rank. To illustrate this, we compare Auto-MFA with EM-based MFA on the first 30 entries in \texttt{iris} dataset. While EM-based MFA fails to run on this dataset, our Auto-MFA converges to a local optimum (loglikelihood = -77, best among 10 random initializations).

\section{Conclusion}

\textbf{GMCM:}
In this paper we present a new algorithm for modeling Gaussian Mixture Copula Models. 
Previous GMCM parameter estimation algorithms, which are based on EM, do not maximize the exact GMCM likelihood nor the algorithms ensure convergence or monotonic increase in likelihood .
We overcome their limitations through the use of Automatic Differentiators. 
Our experiments on a simulated dataset illustrate how our Auto-GMCM can be used to maximize the exact likelihood. 
This work can be extended in many ways. Scalability to high dimensions, theoretical asymptotic properties, such as consistency, also remain to be studied.

\textbf{Mixture of Factor Analyzers:}
In this paper, we show the advantages of using gradient based approach, specifically using exact gradients from automatic differentiation, over the traditional EM-based approach. We discuss its monotonicity and convergence to a better optimal likelihood.  We also demonstrate this applicability to high dimensional MFA, a case where traditional EM-based approach fails. On the face of it, we found that EM-based methods are more robust to random initialization.  Robustness of Auto-MFA vs Traditional EM-based approach to random initialization remains to be studied in detail. 
\bibliographystyle{rss}
\bibliography{example}

\end{document}